\documentclass[aps,prl,twocolumn,showpacs,superscriptaddress]{revtex4} 
\usepackage{graphicx,color} 

\begin{document} 
 
\title{Laser acceleration of ion bunches at the front surface of overdense 
plasmas} 
\author{Andrea Macchi}
\email{macchi@df.unipi.it} 
\affiliation{Istituto Nazionale per la Fisica della Materia (INFM) and
POLYLAB, Dipartimento di Fisica ``E. Fermi'',
Universit\`a di Pisa, Largo B.~Pontecorvo 3, 56127 Pisa, Italy}
\author{Federica Cattani} 
\author{Tatiana V. Liseykina}\thanks{On leave from Institute for Computational 
Technologies, Novosibirsk, Russia} 
\author{Fulvio Cornolti} 
\affiliation{Dipartimento di Fisica ``E. Fermi'' and INFM,
Universit\`a di Pisa, Largo B.~Pontecorvo 3, 56127 Pisa, Italy} 
 
\date{\today} 
 
\begin{abstract} 
The acceleration of ions in the interaction of high intensity laser pulses 
with overdense plasmas is investigated with particle-in-cell simulations. 
For circular polarization of the laser pulses, high-density ion bunches 
moving into the plasma are  
generated at the laser-plasma interaction surface.  
A simple analytical model accounts 
for the numerical observations and provides scaling laws for the  
ion bunch energy and generation time as a function of pulse intensity 
and plasma density.  
\end{abstract} 
 
\pacs{52.38.-r, 52.38.Kd, 52.50.Jm, 52.65.Rr} 
 
\maketitle 
 
The study of the interactions between ultra-intense laser pulses 
and plasmas has proved to be a very rich soil where 
technological progress and fundamental physics meet each other. 
Particularly intriguing is the concept of  
laser-plasma based ion acceleration.  
From astrophysics \cite{astro},  
to medical hadrontherapy \cite{medics},  
from proton radiography \cite{radiography},  
to nuclear physics \cite{nuclear},  
from proton imaging techniques \cite{imaging}, 
to nuclear fusion \cite{fusion}, 
the problem of accelerating and manipulating charged particles with 
laser-plasma interactions offers  
a series of challenges ranging from fundamental to applied physics, 
thus a clear understanding of the basic 
mechanisms is mandatory. 
Several recent experiments have reported the emission of energetic  
ions from solid targets \cite{exp}. 
It is still a matter of debate  
whether the ions are mainly accelerated at the rear surface of the  
target (by the field generated by fast electrons escaping in 
vacuum \cite{mora-betti}) 
or at the front surface involving phenomena such as 
acceleration by a collisionless electrostatic shock 
\cite{denavit,silva,wei}, by a solitary wave \cite{zhidkov} 
or by ion trapping in a propagating double layer \cite{shorokhov}. 
 
In this work we elucidate an even more  
basic process of ion acceleration in cold plasmas, purely related  
to the formation of an electrostatic field due to the action of 
the laser  
ponderomotive force on the electrons and, consequently, on  
the ions via space charge displacement.  
This investigation shows both the necessity of a kinetic description of  
this process and the fundamental role played by the laser light  
polarization by showing the differences between circular and  
linear one. 
It will be shown by particle-in-cell (PIC) simulations  
that circularly polarized light gives rise to  
 a ``pulsed''acceleration and produces ion bunches 
directed into the target. A simple analytical model is used
to explain the acceleration dynamics and 
for the deduction of scaling laws that relate the  
interaction parameters to the energy acquired by the ions. 
With respect to other concepts for laser ion  
acceleration, the present mechanism with circularly polarized light  
leads to  
very high densities in the bunches, as might be of interest  
for problems of compression and acceleration 
of high--density matter. 
 
We consider a laser pulse impinging on  
a cold, step--boundary, overdense plasma with  
$n_{0}/n_{c} =\omega_{p}^2/\omega_L^2> 1$,  
where $n_{0} $ is the initial electron  
density, $n_{c} = m_e \omega_L^2/4\pi e^2$ 
is the critical density for  
a laser with carrier frequency $\omega_L$, 
$\omega_p$ is the plasma frequency and $m_e$, $e$ are the  
electron mass and charge. The laser field  
amplitude will be given 
in units of the dimensionless parameter 
$a_L=(eE_L/m_e\omega_L c)$. 
In the PIC simulations, the resolution is high enough to resolve  
low density regions, sharp gradients and the dynamics of both electrons and  
ions by taking at least 20 particles per cell at $n_e=n_c$ and a  
spatial resolution  
better than $0.1 d_p$ where $d_p=c/\omega_{p}$. 
For reference, in all simulations shown 
the target boundary is located at $x=0$ and
the laser impinges on it at $t=0$. 

\begin{figure}
\begin{center}
\includegraphics[width=8.5cm]{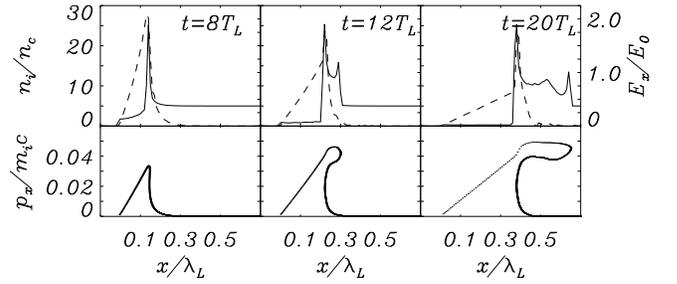}
\end{center}
\caption{Ion density $n_i$ (top, solid line) 
electrostatic field $E_x$ (top, dotted), and ($x,p_x$) phase space  
projection (bottom) at different times (labels)  
from 1D PIC simulations. Run parameters are $a_L=2$, $n_e/n_c=5$, 
$\ell=2\lambda_L$. $E_0=m_e\omega c/e$. 
} \label{fig:bunch}  
\end{figure} 
 
The essential features of the generation of ion bunches can be seen  
from one--dimensional (1D) simulations. 
The laser pulse is circularly polarized,  
incident from the left side and propagating along the $x$ 
axis; its amplitude rises for $6T_L$ 
(where $T_L=2\pi/\omega_L$ is the laser period) 
up to the value $a_L = 2$ 
and then remains constant. 
The plasma has a slab profile with  
$n_{0} = 5 n_{c}$ 
and thickness $\ell=2\lambda_L$. 
Electrons and ions are  
assumed to be cold, $T_e = T_i = 0$, 
and ion charge and mass number are $Z = 1$, $A=1$. 

Fig.~\ref{fig:bunch} 
shows the ion density profiles and ($x,p_x$) phase space 
projections at different times.  
At $t=8T_L$ a sharp peak of the ion density 
reaching up to seven times the initial value is shown which moves  
inwards the plasma with an average velocity of about $0.013c$  
leaving behind a low--density shelf with velocity linearly  
decreasing. At $t=12T_L$,  
after the density at the peak has reached the maximum 
value of about $7n_{0}$ and the velocity has reached the value 
of $0.02c$, 
a second density peak has appeared
on the right of the first peak and moves into the plasma 
at the speed of about $0.04c$ with  
the ($x,p_x$) distribution now clearly  
a two--valued function of $x$. 

\begin{figure}
\begin{center}
\includegraphics[width=8.0cm]{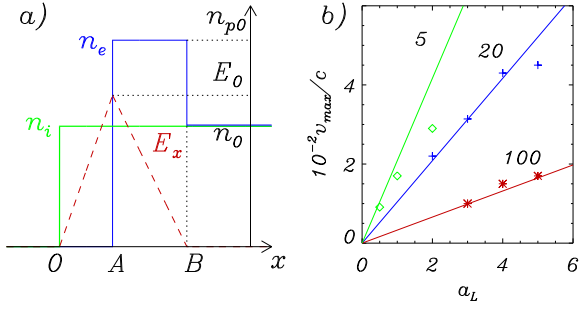} 
\end{center}
\caption{(Color online) 
a): Schematic of the model profiles of the ion density (green, thick) 
the electron density (blue, thin) 
and the electric field (red, dashed) at the initial stage
when the electrons are in equilibrium and the ions have not 
moved yet. The label $A$ indicates  
the electron front where the laser evanescence starts, $B$ indicates  
the point where the ponderomotive force 
vanishes so that $\overline{AB} = l_s$.
b): Comparison between the ion bunch velocity observed in simulations 
(data points) and the model prediction (lines), 
as a function of the laser pulse amplitude $a_L$ and the normalized density 
$N=n_{i0}/n_c$. The cases studied are:  
$N=100$, $Z/A=1/2$ (red, stars); 
$N=20$, $Z/A=1$ (blue, crosses); 
$N=5$, $Z/A=1/2$ (green, diamonds).} 
\label{Fig:scheme} \label{fig:scaling}
\end{figure} 

A simple model for this ion bunch generation  
can be inferred. 
When the laser pulse impinges on the plasma surface, 
electrons are quickly pushed inward by the 
ponderomotive force, 
 i.e. the steady part of the 
${\bf v}\times{\bf B}$ force. (Notice that the
oscillating part of the ${\bf v}\times{\bf B}$ force is zero
for circular polarization; this is the reason why, 
as shown below, the interaction regime is completely
different for linear polarization).   
The electrons pile up leaving behind a charge depletion layer  
and giving rise to an electrostatic field $E_x$  
back--holding them. 
We assume that they quickly reach an equilibrium position where $E_x$  
balances the ponderomotive force exactly.  
Let us take for simplicity a linear profile of $E_x$ 
both in the depletion layer ($E_x=E_{0}x/d$ for $0<x<d$)  
and in the compression region  
($E_x=E_0[1-(x-d)/l_s]$ for $d<x<d+l_s$), which implies 
 a uniform electron density $n_{p0}$  
in this region, see Fig.~\ref{Fig:scheme}. 
The parameters $E_0$, $n_{p0}$, 
$d$ and $l_s$ are related by the equations 
$E_0=4\pi e n_{0}d$ (due to Poisson equation),  
$n_{0}(d+l_s)=n_{p0}l_s$ (due to global charge conservation),  
and $E_{0}e n_{0} l_s/2 \simeq 2I_L/c$
(due to the balance between the total radiation and electrostatic
pressures).

The electrostatic field on the ions starting at initial positions $x_0<d$  
is a constant over the trajectories of the ions, $E_x=E_x(x_0)$, and  
increases with $x_0$. Thus, these ions will never reach 
those with $x_0>d$, the ion density will decrease and a ion ``shelf'' 
will be formed, the field on the leftmost ion layer being zero.  
As for the ions with initial position in the compression region  
$d<x_0<d+l_s$, assuming that the electrostatic field is a function 
of $x_0$ for these ions too is consistent with the  
assumption that the electrons follow a quasi--equilibrium dynamics
and the ponderomotive force equals the electrostatic field at any time, 
the total pressure being always $2I_L/c$.  
Being the electrostatic field a linear function of 
$x_0$, all ions will reach the point $B$ at the same 
time and the ion density will assume an infinite value there, i.e. 
the hydrodynamic description breaks down. The ion bunch is thus due 
to accelerated ions that cross the point $B$ and are injected into the 
unperturbed plasma region, where they can move ballistically provided 
that the charge unbalance is neutralized by electrons  
accompanying the bunch. 
 The total number of accelerated ions per unit surface is 
$n_{p0}l_s$. 

Note that the model assumes the point $B$, where the field vanishes, 
not to move during compression, and
predicts ions to have a a flat--top velocity distribution 
extending between $0$ and $v_{max}$ and thus the density to be uniform
in the region behind the ion front. 
Actually, since for the evanescence length $l_s$ we expect
$l_s \simeq d_p \sim n_e^{-1/2}$,  
during the compression of the ion fluid 
the field will tend to penetrate 
deeper into the plasma keeping the field at the 
surface and the total electrostatic pressure constant. 
Thus, ions beyond point $A$ will be accelerated by a field decreasing  
in time and will get to the breaking point later. 
This effect causes the ion bunch to be more localized both in coordinate
and velocity space.

Our model thus gives a scenario in qualitative  
agreement with the numerical observations and also provides 
quantitative estimates and scaling laws. 
Denoting as $x_0=\zeta_0+d$ the initial position of an ion  
with mass $m_i$  
and charge $q_i$ in the compression region, ($0<\zeta_0<l_s$), 
the force acting on this ion is given by 
$ F_i = q_i E_0(1 -{\zeta_0}/{l_{s}})$.  
Thus, The velocity of an initially immobile ion
is $v_i=({q_i E_0}/{m_i})(1-{\zeta_0}/{l_{s}})t$  
and the position is  
$x_i=({q_i E_0}/{2m_i})(1-{\zeta_0}/{l_{s}})t^2+x_0$. 
The ``breaking'' time at which all ions reach point $B$ is  
$\tau_i = \sqrt{{2 l_{s} m_i}/{q_i E_0}}$. 
The maximum velocity 
is thus $v_{max} = \sqrt{{2l_{s}q_{i}E_{0}}/{m_i}}=2v_a$, 
being $v_a={l_{s}}/{\tau_i}$ the average ion velocity, i.e.
the velocity of the laser front.
By relating the model parameters 
to the laser intensity and plasma density  
we obtain 
\begin{equation}\label{eq:vscaling} 
\frac{v_a}{c} = \sqrt{\frac{Z}{A}\frac{m_e}{m_p}\frac{n_c}{n_e}}a_L 
\;\; {\mathrm{and}} \;\; 
\tau_i \simeq \frac{1}{\omega_{L}a_L}\sqrt{\frac{A}{Z}\frac{m_p}{m_e}} 
\end{equation} 
where $l_s \simeq d_p$ has been assumed. 
The predicted velocity scaling has been tested  
via numerical simulations, yielding a rather good agreement  
with the model, see Fig. \ref{fig:scaling}. 
For instance, for the parameters of Fig.~\ref{fig:bunch}  
the model predicts $v_a/c = 0.021$ and $\tau_i \simeq 3.4T_L$;
in the simulation the breaking of the ion 
front occurs at $t \approx 10T_L$ which, taking into account
the pulse rise time of $6T_L$, is in fair
agreement with the model prediction. 

At the ``breaking'' of the ion profile with the formation of the 
bunch, the equilibrium between the electrostatic and 
the ponderomotive forces on electrons 
is lost and, if the laser pulse is not over, electrons rearrange  
themselves to provide a new equilibrium. The process of bunch formation  
might then restart although with different initial conditions,  
so that the simple model is not adequate anymore for quantitative predictions. 
In Fig.~\ref{fig:bunch}
a second wide bunch actually appears at 
$t \simeq 15T_L$ and then propagates into the plasma. 
At later times, bunches with shorter duration are  
generated at a higher rate. 

\begin{figure}
\begin{center}
\includegraphics[width=8.5cm]{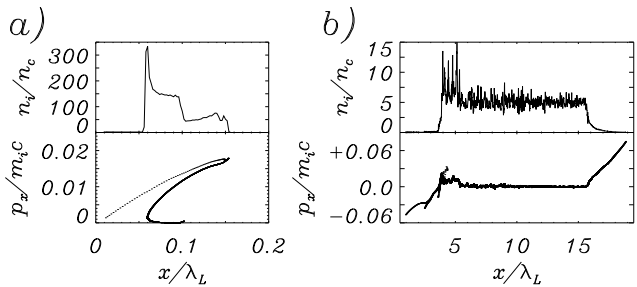}
\end{center} 
\caption{a) Ion bunch acceleration from a thin ($\ell=0.1\lambda_L$), 
solid--density ($n_0/n_c=100$) target irradiated by a short laser pulse with  
amplitude $a=5$ and duration of $6T_L$ (FWHM). 
The ion density profile (thick line) and the 
($x,p_x$) phase space distribution (dotted) are shown at $t=13T_L$.
b) Ion density (top) and
phase space ($x,p_x$) projection (bottom)  
at $t=86T_L$, for the same simulation 
parameters of Fig.~\ref{fig:bunch},  
but linear polarization of the laser pulse.} 
\label{fig:LP}  
\label{fig:thin} 
\end{figure} 

For potential applications, 
it is interesting to study the possibility to produce a single bunch  
from a thin solid target. 
Results for the case of an $n_0/n_c=100$, $Z/A=1/2$ target  
with a thickness $\ell=0.1\lambda_L$ and a short pulse 
with intensity $a_L=5$ 
are shown in Fig.~\ref{fig:thin}.  
The temporal envelope for the field is given by 
$\sin^2(\pi t/\tau)$ with $\tau=12$. 
A single bunch of ions with velocity $\approx 0.017c$ 
is produced at $t \approx 5.5T_L$ 
and leaves the target from the back  
side. The ion bunch charge is completely neutralized by electrons 
both inside and outside the target. 

Collisional ion stopping (not included in the PIC code)
should not affect ion bunch generation strongly 
for thin, low--$Z$ targets.
In fact, taking for the example of Fig.\ref{fig:thin}
$\lambda=0.5~\mu\mbox{m}$, so that $n_0=4\times 10^{23}~\mbox{cm}^{-3}$, 
the stopping power in the cold plasma given by the Bohr formula
\cite{atzeni} is $\approx 0.19Z^2~\mbox{MeV}/\mu\mbox{m}$ for ions
with $v=0.017c$ and charge $Z=A/2$, and 
the stopping length is thus $\approx 1.4Z^{-1}~\mu\mbox{m}$, 
lower than the target thickness in the example if $Z<28$.
For lower densities, typical of preformed
plasmas or, e.g., foam targets, collisional ion stopping is even less
important. 
Collective ion stopping in the plasma is
included in the PIC simulation, and is found not to affect ion transport
in the regimes addressed by our simulations.

The mechanism of ion bunch formation outlined above works cleanly 
for circular laser polarization. For linear polarization
the oscillating part of the ${\bf v}\times{\bf B}$ force causes 
strong electron heating (up to MeV energies)
which changes the regime of ion acceleration 
qualitatively. 
Fig.~\ref{fig:LP} shows results from 
a simulation with parameters identical to Fig.~\ref{fig:bunch}
but the laser polarization that was changed to linear.  
No clear propagating ion bunch as that of Fig.~\ref{fig:bunch}
is observed. The hot 
electrons form a Debye sheath leading to ion acceleration towards 
vacuum at the back surface as soon as they reach it while a few 
ions are also accelerated  
from the front surface towards vacuum in a similar way
(Fig.~\ref{fig:LP}).
Near the laser pulse front, the phase space structures  
look similar 
to those generated by the shock acceleration investigated in  
Ref.\cite{silva} at higher intensity. For the present parameters, 
sheath acceleration dominates at the time shown.  
In contrast, for circular polarization, it is found that 
most of the electrons are heated up  
to energies of a few keV behind the ion bunch and at the  
laser--plasma interaction surface. 
 
Notice that in the case of shock acceleration 
(see e.g. Ref.\cite{silva}) the laser pulse acting as a 
piston drives a shock wave into the plasma which reflects the  
ions thus accelerated up 
to velocities twice the shock speed. 
In the present case the fastest  
ions have twice the piston velocity, i.e. the velocity of the  
laser reflection front at the breaking time, and they come from 
\emph{behind} the front. It is also worth to notice that the present 
mechanism of ion bunch formation is of
electrostatic and kinetic nature while a purely hydrodynamic 
description is not adequate. 

\begin{figure}
\begin{center}
\includegraphics[width=8.5cm]{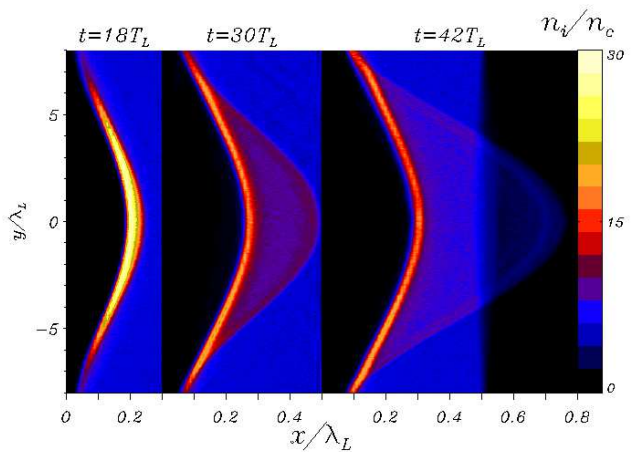} 
\end{center}
\caption{(Color online) Contours of ion density $n_i/n_c$ 
from 2D simulations at three different times.
The laser pulse
impinges from the left and has a transverse FWHM of
$4\lambda_L$. Notice the different scales on the $x$ and $y$ axis.} 
\label{fig:2D} 
\end{figure} 

We performed 2D simulations to show that
the mechanism for bunch formation is also at play in more than one dimension, 
and in particular in the case of a laser pulse with a finite transverse 
width that bends the plasma surface (hole boring). 
Fig.\ref{fig:2D} shows results for a case where the laser pulse 
has a Gaussian intensity profile in the transverse ($y$) direction 
with halfwidth $w = 4 \lambda_L$, duration $\tau = 12T_L$, and
peak intensity $a_L=2$.
The target thickness is $\ell=0.5\lambda_L$ and the density is
$n_{i0}=5n_c$.
 At $t=18T_L$, we see the gaussian--shape compression
front produced by hole boring. At $t=30T_L$, the ion bunch has 
formed and propagates into the plasma, keeping the shape of
the compression front. At $t=42T_L$, the bunch ions in the central 
region have left the target and propagate in vacuum.
Another run was performed with the same 
parameters but a ``tighter'' pulse, i.e. $w=2\lambda_L$.
and almost identical results were found.

\begin{figure}
\begin{center}
\includegraphics[width=8.5cm]{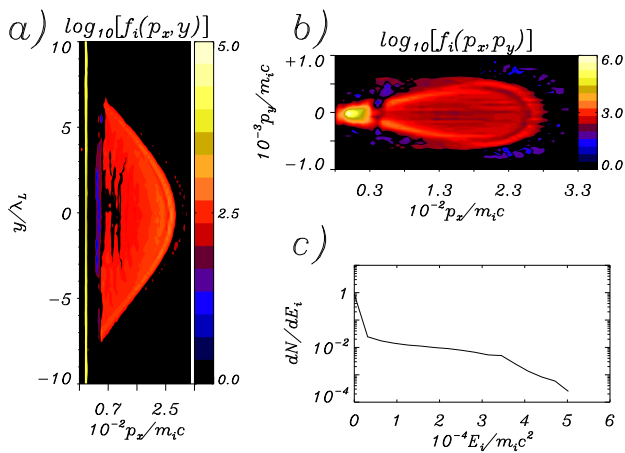} 
\end{center}
\caption{(Color online) The $(y,p_x)$ and $(p_x,p_y)$ phase
space projections [a)-b)] and the energy spectrum [c)]
of ions from the 2D simulation of Fig.~\ref{fig:2D},
at $t=47T_L$.}  
\label{fig:2Dps} 
\end{figure} 

Fig.\ref{fig:2Dps} shows the ion phase space projections and 
the energy spectrum
for the simulation of Fig.\ref{fig:2D}, at $t=47T_L$.
The dependence of the longitudinal momentum $p_x$ on $y$ 
[Fig.\ref{fig:2Dps} a)] and the energy spectrum
[Fig.\ref{fig:2Dps} c)] 
resemble the intensity modulation of the laser pulse.
The relative spread in the transverse momentum [Fig.\ref{fig:2Dps} b)]
is very small, 
$({\langle p_y^2\rangle/\langle p_x^2\rangle})^{1/2} \simeq2\times 10^{-2}$, 
showing that also in 2D the acceleration is perpendicular to the 
target surface.
In the $w=2\lambda_L$ case we find slightly slower ion velocities and
almost the same beam divergence.

In some runs, some ``rippling'' of the ion density front is observed. 
This rippling appears to be due to electron oscillations at the surface 
(around their quasi--equilibrium position) excited due to
the finite rise time and the curvature of the laser pulse.  
The rippling is ``imprinted'' in the ion density
at later times, but has no noticeable effect on the dynamics of
ion acceleration. We conclude that 2D effects do not 
qualitatively affect ion bunch formation. 

In conclusion, a regime of laser ion acceleration in cold overdense plasmas
by circularly polarized pulses has been characterized by means of 
PIC simulations and a simple model. The dynamics is qualitatively 
different from the case of linear polarization where ion acceleration is
driven by fast electrons. With the present scheme one may obtain moderate
ion energies but very high ion densities and low beam divergence.

We acknowledge useful discussions with S. Atzeni and F. Pegoraro.
Part of the numerical simulations were performed on the Linux cluster at  
the CINECA facility (Bologna, Italy), with the support 
of the INFM supercomputing initiative.

\end{document}